\documentclass{llncs}
\usepackage[usenames]{color}

\usepackage{qtree}
\usepackage[dvips]{graphicx}
\usepackage{gastex}

\usepackage{amsmath}
\usepackage{amssymb}
\usepackage{amsfonts}
\usepackage{amscd}
\usepackage{amsbsy}

\usepackage{verbatim}
\usepackage[T1]{fontenc}
\usepackage[latin1]{inputenc}
\usepackage{enumerate}

\definecolor{Gray}{rgb}{0.9,0.9,0.9}
\definecolor{Gray2}{rgb}{0.6,0.6,0.6}

\usepackage{subfigure}

\usepackage{latexsym}

\newcommand{\A}{\mathcal{A}}
\newcommand{\B}{\mathcal{B}}
\newcommand{\F}{\mathcal{F}}
\newcommand{\G}{\mathcal{G}}
\newcommand{\N}{\mathcal{N}}
\newcommand{\T}{\mathcal{T}}

\newcommand{\TT}{\mathfrak{T}}

\DeclareMathOperator{\Sh}{\mathsf{X}}

\begin{document}
\title{Cellular automata on regular rooted trees}
\titlerunning{Cellular automata on regular rooted trees}
\author{Tullio Ceccherini-Silberstein\inst{1}\and~Michel Coornaert\inst{2}\and~Francesca Fiorenzi\inst{3}\and~Zoran \v Suni\'c\inst{4}
}
\authorrunning{T. Ceccherini-Silberstein, M. Coornaert, F. Fiorenzi and Z. \v Suni\'c}
\institute{Dipartimento di Ingegneria, Universit\`a del Sannio\\
C.so Garibaldi 107, 82100 Benevento, Italy\\
\email{tceccher@mat.uniroma3.it}
\and
Institut de Recherche Math\'ematique Avanc\'ee,
UMR 7501, Univ. Strasbourg\\
7 rue Ren\'e Descartes, 67084 Strasbourg Cedex, France\\
\email{coornaert@math.unistra.fr}
\and
Laboratoire de Recherche en Informatique, UMR 8623, Bât 650 Univ. Paris-Sud 11\\
91405 Orsay Cedex France 91405 Orsay, France\\
\email{fiorenzi@lri.fr}
\and
Department of Mathematics, Texas A\&M University\\
 MS-3368, College Station, TX 77843-3368, USA\\
\email{sunic@math.tamu.edu}
}

\maketitle

\begin{abstract}
We study cellular automata on regular rooted trees. This includes the characterization of sofic tree shifts in terms of unrestricted Rabin automata and the decidability of the surjectivity problem for
cellular automata between sofic tree shifts.
\end{abstract}
%\subjclass[2000]{03B25, 05C05, 37B10, 37B15, 68Q70, 68Q80}

\section{Introduction}
In this paper, we study cellular automata between subshifts of $A^{\Sigma^*}$ (also called tree shifts), where $A$ is a finite nonempty set and $\Sigma^*$ is a finitely generated free monoid identified with the $\vert\Sigma\vert$-regular rooted tree. We investigate, in particular, the decidability of the surjectivity problem for these cellular automata.

Amoroso and
Patt~\cite{AmorosoPatt72} proved that the surjectivity and the
injectivity problems have a positive answer in the
one-dimensional case (i.e. when $\Sigma^*$ is replaced by $\mathbb{Z}$ or $\mathbb{N}$). On the other hand,
Kari~\cite{Kari94} proved that these problems fail to
be decidable in dimension $d\geq2$ (i.e. for cellular automata defined on $A^{\mathbb{Z}^d}$).
There are more general algorithms to decide the surjectivity of a cellular automaton on a finitely generated free monoid
(which are deducible combining the results of Rabin~\cite{Rabin69}, Muller and Schupp~\cite{MullerSchupp85} and Thatcher and Wright~\cite{ThatcherWright68}), but they are not practical. We have worked out the details in a limited setting of interest, namely when we start with an unrestricted Rabin automaton.

Tree shifts have been extensively studied by Aubrun and Béal in~\cite{Aubrun11}, \cite{AubrunBeal10} and \cite{AubrunBeal12}.
In the present work we use a slightly different (but equivalent) setting.

A tree shift is said to be of finite type if it can be described as the set of configurations avoiding a finite number of forbidden patterns.
Sofic tree
shifts are defined as the images of tree shifts of finite type under
cellular automata.
In the one-dimensional case, a sofic subshift of $A^\mathbb{N}$ may be
characterized as the set of all right-infinite words accepted by some
finite-state automaton. In our setting, we use the notion of an unrestricted Rabin
automaton (see \cite{Rabin69}, \cite{ThatcherWright68}, \cite{tata07}), as well as a related notion of acceptance, in order to provide
the analogous characterization of sofic tree shifts.

Let us now illustrate our decidability results. It is easy to decide the emptiness of a sofic tree shift accepted by a given unrestricted Rabin automaton. An idea to decide the surjectivity of a cellular automaton $\tau \colon A^{\Sigma^*}\to A^{\Sigma^*}$ could be to establish the emptiness of the tree language $A^{\Sigma^*} \setminus \tau(A^{\Sigma^*})$.
But given a nontrivial tree shift, its complement
always fails to be a subshift (nevertheless, Rabin theory guarantees that this tree language is still recognizable by a general Rabin automaton). In order to avoid this obstacle, we introduce the set of full-tree-patterns of a tree shift which is a finite tree language that characterizes the shift. Moreover, the full-tree-patterns of a sofic tree shift are recognizable by a suitably defined finite-tree automaton.

We prove that the recognizable sets of full-tree-patterns form a class which is closed under complementation and for which the emptiness problem is decidable.
This allows us to find algorithms establishing both the surjectivity of a cellular automaton $\tau \colon A^{\Sigma^*}\to A^{\Sigma^*}$ and the equality of two sofic-tree shifts presented by unrestricted Rabin automata. With this latter result at hand we can provide a general algorithm establishing the surjectivity of a cellular automaton defined between sofic tree shifts.

In this paper we just detail the decision procedures we mentioned above. The proofs of the results leading to these algorithms are omitted.

\section{Definitions and background}
In the sequel, we denote by $\Sigma$ and $A$ two nonempty finite sets. In particular, the set $A$ is called \emph{alphabet} and its elements are called \emph{labels} or \emph{colors}.

\subsection{The free monoid $\Sigma^*$}\label{ss;free}
For $n \in \mathbb{N}$, we denote by $\Sigma^n$ the set of all \emph{words} $w = \sigma_1\sigma_2 \cdots \sigma_n$ of \emph{length} $n$ (where $\sigma_i \in \Sigma$ for $i=1,2,\dots,n$) over $\Sigma$. In particular $\varepsilon \in \Sigma^0$ indicates the only word of length $0$ called the \emph{empty word}. For $n \geq 1$, we denote by $\Delta_n$ the set $\bigcup_{i = 0}^{n-1}\Sigma^i$ (that is, the set of all words of length $\leq n-1$).

The \emph{concatenation} of two words $w = \sigma_1\sigma_2 \cdots \sigma_n \in \Sigma^n$ and $w' = \sigma'_1\sigma'_2 \cdots \sigma'_m\in \Sigma^m$ is the word $ww' = \sigma_1\sigma_2 \cdots \sigma_n\sigma'_1\sigma'_2 \cdots \sigma'_m\in \Sigma^{m+n}$. Then the set $\Sigma^* = \bigcup_{n \in \mathbb{N}} \Sigma^n$, equipped with the multiplication given by concatenation, is a monoid with identity element the empty word $\varepsilon$. It is called the \emph{free monoid} over $\Sigma$.

From the graph theoretical point of view, $\Sigma^*$ is the vertex set of the $\vert \Sigma\vert$-regular rooted tree. The empty word $\varepsilon$ is its root and, for every vertex $w \in \Sigma^*$, the vertices $w\sigma\in \Sigma^*$ (with $\sigma \in \Sigma$) are called the \emph{children} of $w$. Each vertex is connected by an edge to each of its children.

\subsection{Configurations and tree shifts}
We denote by $A^{\Sigma^*}$ the set of
all maps $f \colon  \Sigma^* \to  A$. It is called the \emph{space of configurations} of $\Sigma^*$ over the alphabet $A$. When equipped with the \emph{prodiscrete topology} (that is, with the product topology  where each factor $A$ of $A^{\Sigma^*} = \prod_{w \in \Sigma^*}A$ is endowed with the discrete topology), the configuration space is a compact, totally disconnected, metrizable space.
Also, the free monoid $\Sigma^*$ has a right action on $A^{\Sigma^*}$ defined as follows: for every $w \in \Sigma^*$ and $f \in  A^{\Sigma^*}$ the configuration $f^w \in  A^{\Sigma^*}$ is defined by setting
$f^w(w') = f(ww')$ for all $w' \in \Sigma^*$. This action, called the \emph{shift action}, is continuous with respect to the prodiscrete topology.

Recall that
a neighborhood basis of a configuration $f \in A^{\Sigma^*}$ is given by the sets
$\N(f,n) = \{g \in A^{\Sigma^*} : g\vert_{\Delta_n} = f\vert_{\Delta_n}\}$
where $n \geq 1$ (as usual, for $M\subset \Sigma^*$, we denote by $f\vert_M$ the restriction of $f$ to $M$).

A subset $X \subset A^{\Sigma^*}$ is called a \emph{subshift} (or \emph{tree shift}, or simply \emph{shift}) provided that $X$ is closed (with
respect to the prodiscrete topology) and \emph{shift-invariant} (that is, $f^w \in X$ for all
$f \in X$ and $w \in \Sigma^*$).

\subsection{Forbidden blocks and shifts of finite type}

Let $M\subset \Sigma^*$ be a finite set. A \emph{pattern} is a map $p \colon  M \to  A$. The set $M$ is called the \emph{support} of $p$ and it is denoted by ${\rm supp}(p)$. We denote by $A^M$ the set of all patterns with support $M$.
A \emph{block} is a pattern $p \colon  \Delta_n \to  A$. The integer $n$ is called the \emph{size} of the block.
The set of all blocks is denoted by $\B(A^{\Sigma^*})$.

If $X$ is a subset of $A^{\Sigma^*}$ and $M \subset \Sigma^*$ is finite,
the set of patterns $\{f\vert_M : f \in X\}$ is denoted by $X_M$. For $n \geq 1$, the notation $X_n$ is an abbreviation for $X_{\Delta_n}$ (that is, the set of all blocks of size $n$ which are restrictions to $\Delta_n$ of some configuration in $X$). We denote by $\B(X)$ the set of all blocks of $X$ (that is, $\B(X)= \bigcup_{n\geq1}X_n$).

Given a block $p \in \B(A^{\Sigma^*})$ and a configuration $f \in A^{\Sigma^*}$, we say that $p$ \emph{appears} in $f$ if there exists $w \in \Sigma^*$ such that $(f^w)\vert_{{\rm supp}(p)} = p$. If $p$ does not appear in $f$, we say that $f$ \emph{avoids} $p$.
Let $\F$ be a set of blocks. We denote by $\mathsf{X}(\F)$ the set of all configurations in $A^{\Sigma^*}$ avoiding each block of $\F$, in symbols
$\mathsf{X}(\F) = \{f \in A^{\Sigma^*} : (f^w)\vert_{\Delta_n} \notin \F, \textup{ for all } w \in \Sigma^* \textup{ and } n \geq 1 \}.$

In analogy with the one-dimensional case (see for example \cite[Theorem 6.1.21]{LindMarcus95}), we have the following combinatorial characterization of subshifts: \emph{
a subset $X  \subset A^{\Sigma^*}$ is a subshift if and only if there exists a set
$\F \subset \B(A^{\Sigma^*})$ of blocks such that
$X = \mathsf{X}(\F)$.}

Let $X \subset A^{\Sigma^*}$ be a subshift. A set $\F$ of blocks as above is called a \emph{defining set of forbidden blocks} for $X$.
A subshift is \emph{of finite type} if it admits a finite defining set of forbidden blocks.

\begin{remark}\label{memoryREM}
We can always suppose that the forbidden blocks of a defining set of a given subshift of finite type all have the same support. This motivates the following definition: a shift of finite type has \emph{memory} $n$ if it admits a defining set of forbidden blocks of size $n$.
Notice that a shift with memory $n$, also has memory $m$ for all $m\geq n$.
\end{remark}

\begin{figure}[!h]
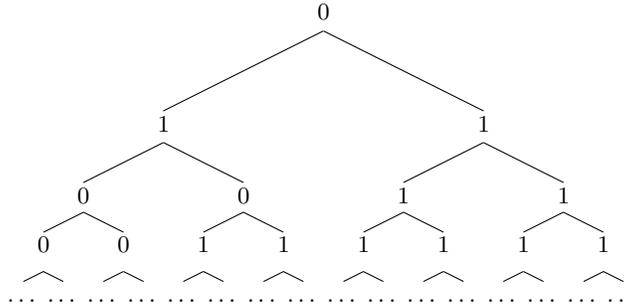

%%%%%%%%%%%%%%%%%%%%%%%%%%%%%%%%%%%%%%%%%%%%%%%%%%%%%%%%%%%%%%%%%%%%%%%
\qtreecentertrue
\Tree [.$0$ [.$1$ [.$0$ [.$0$ $\dots$ $\dots$ ] [.$0$ $\dots$ $\dots$ ] ] [.$0$ [.$1$ $\dots$ $\dots$ ] [.$1$ $\dots$ $\dots$ ] ] ] [.$1$ [.$1$ [.$1$ $\dots$ $\dots$ ] [.$1$ $\dots$ $\dots$ ] ] [.$1$ [.$1$ $\dots$ $\dots$ ] [.$1$ $\dots$ $\dots$ ] ] ] ]
%%%%%%%%%%%%%%%%%%%%%%%%%%%%%%%%%%%%%%%%%%%%%%%%%%%%%%%%%%%%%%%%%%%%%%%
\caption{A configuration of the tree shift presented in Example~\ref{e:monochromaticchildren}.}\label{ALBmonochromaticchildren}
\end{figure}

\begin{example}[Monochromatic children]\label{e:monochromaticchildren}
Since $\Delta_2 = \{\varepsilon\} \cup \Sigma$, we can identify $A^{\Delta_2}$ with $A \times A^\Sigma$. Consider the set of blocks $$\F = \left\{(a,(a_\sigma)_{\sigma \in \Sigma}) \in A \times A^\Sigma  : a_{\sigma} \neq a_{\sigma'} \textup{ for some } \sigma,\sigma' \in \Sigma\right\}.$$ The tree shift $\Sh(\F) \subset A^{\Sigma^*}$ is of finite type and exactly consists of those configurations for which every vertex in $\Sigma^*$ has monochromatic children.
If $\vert \Sigma \vert = 2$ and $A = \{0,1\}$ an example of a configuration in $\Sh(\F)$ is given in Figure~\ref{ALBmonochromaticchildren}.
\end{example}

\subsection{Cellular automata and sofic tree shifts}
Let $X \subset A^{\Sigma^*}$ be a tree shift.
A map $\tau \colon   X \to  A^{\Sigma^*}$ is called a \emph{cellular automaton} if it satisfies the following condition: there exists a finite subset $M \subset \Sigma^*$ and a map $\mu \colon  A^M \to  A$ such that $\tau(f)(w) = \mu((f^w)\vert_M)$ for all $f \in X$ and $w \in \Sigma^*$. The set $M$ is called a \emph{memory set} for $\tau$ and $\mu$ is the associated \emph{local defining map}.
We assume in the sequel (without loss of generality), that a memory set has the form $M = \Delta_n$, for a suitable $n\geq 1$.

The Curtis-Hedlund-Lyndon theorem gives a topologically characterization of cellular automata: \emph{a map $\tau \colon   X \to  A^{\Sigma^*}$ is a cellular automaton if and only if it commutes with the shift action (that is, $(\tau(f))^w = \tau(f^w)$ for all $f \in X$ and $w \in \Sigma^*$), and is
continuous (with respect to the prodiscrete topology on $X$)}.
 For a proof in the one-dimensional case, see \cite[Theorem 6.2.9]{LindMarcus95}. See also \cite[Theorem 1.8.1]{livre} and \cite{Fiorenzi03}, for a more general setting.
It immediately follows that the image of a tree shift under a cellular automaton is still a tree shift.

\begin{remark}\label{r:samealphabet}
In the definition of a cellular automaton we have assumed that the alphabet of the shift $X$ is the same as the alphabet of its image $\tau(X)$. In this assumption there is no loss of generality because if $\tau \colon X \to B^{\Sigma^*}$, one can always consider $X$ as a subshifts of $(A \cup B)^{\Sigma^*}$.
Classically, a cellular automaton is also a selfmapping $\tau \colon X \to X$. By dropping this hypothesis, we deal with a more general notion that, in the one-dimensional case, corresponds to that of \emph{sliding block code} as defined in~\cite{LindMarcus95}.
\end{remark}

A subshift $X \subset A^{\Sigma^*}$ is called \emph{sofic} provided there
exist a subshift of finite type $Y \subset A^{\Sigma^*}$ and a cellular automaton $\tau \colon   Y \to  A^{\Sigma^*}$
such that $X = \tau(Y)$.

\begin{remark}
Every subshift of finite type is sofic but there are examples of sofic subshifts which are not of finite type (see \cite[Example 2.1.5, Example 2.1.9]{LindMarcus95}).
\end{remark}

\section{Unrestricted Rabin graphs and automata}\label{Unrestricted Rabin automata}
An \emph{unrestricted Rabin graph}, is a $4$-tuple $\G = (S,\Sigma,A,\T)$, where
$S$ is a nonempty set, called the set of \emph{states} (or \emph{vertices}) of $\G$ and $\T$ is a subset of $S \times A \times S^\Sigma$ whose elements are called \emph{transition bundles}.
When the state set $S$ is finite $\G = (S,\Sigma,A,\T)$ is called an \emph{unrestricted Rabin automaton}.

Given a transition bundle $t = (s;a;(s_\sigma)_{\sigma \in \Sigma}) \in \T$ we denote by
$\mathbf{i}(t) := s \in S$ its \emph{initial state}, by $\lambda(t) := a \in A$ its \emph{label}, by $\mathbf{t}(t) := (s_\sigma)_{\sigma \in \Sigma} \in S^\Sigma$ its \emph{terminal sequence} and by $\mathbf{t}_\sigma(t) :=  s_\sigma \in S$ its $\sigma$-\emph{terminal state}.
A \emph{bundle loop on $s \in S$} is a transition bundle $t \in \T$ such that $\mathbf{i}(t) = \mathbf{t}_\sigma(t) = s$ for all $\sigma \in \Sigma$.

An unrestricted Rabin graph $\G = (S,\Sigma,A,\T)$ is said to be \emph{essential} provided that for each state $s \in S$ there is a transition bundle starting at $s$.

\begin{definition}[Unrestricted Rabin graph of a configuration]
\label{e:configuration}
{\rm The \emph{unrestricted Rabin graph of a configuration $f \in A^{\Sigma^*}$} is defined by $\G_f = (\Sigma^*,\Sigma,A,\T_f)$
where
$\T_f = \{(w;f(w);(w\sigma)_{\sigma \in \Sigma}) : w \in \Sigma^*\}.$
}\end{definition}

\begin{definition}[Homomorphism]\label{homomorphism}
{\rm A \emph{homomorphism} from $\G_1 = (S_1,\Sigma,A,\T_1)$ to $\G_2 = (S_2,\Sigma,A,\T_2)$ is a map $\alpha \colon  S_1 \to  S_2$ such that
$(\alpha(s);a;(\alpha(s_\sigma))_{\sigma\in\Sigma}) \in \T_2$
for all $(s;a;(s_\sigma)_{\sigma\in\Sigma}) \in \T_1$.
By abuse of language/notation, we also denote by $\alpha \colon  \G_1 \to  \G_2$ such a
homomorphism.
}\end{definition}

\begin{definition}[Acceptance]
{\rm Let $\A = (S,\Sigma,A,\T)$ be an unrestricted Rabin automaton. We say that a configuration $f \in A^{\Sigma^*}$ is \emph{accepted} (or \emph{recognized}) by $\A$, if there exists a homomorphism $\alpha \colon  \G_f \to  \A$.
In this case, we say that \emph{$f$ is accepted by $\A$ via} $\alpha$.
We denote by $\Sh_\A$ the set consisting of all those configurations $f \in A^{\Sigma^*}$ accepted by $\A$.
An unrestricted Rabin automaton $\A$ is called a \emph{presentation} for $X \subset A^{\Sigma^*}$ provided that $X = \Sh_\A$.
}\end{definition}

\begin{remark}
In the sequel, we shall always consider essential unrestricted Rabin automata. This is not restrictive since, by recursively removing all states that are source of no transition bundles, we can transform any unrestricted Rabin automaton $\A$ into an essential one
$\A'$ which accepts the same subset, i.e. such that $\Sh_\A = \Sh_{\A'}$.
\end{remark}

\begin{remark}\label{r:acceptance} Explicitly, a configuration $f \in A^{\Sigma^*}$ is accepted by an unrestricted Rabin automaton $\A = (S,\Sigma,A,\T)$ if there exists a map $\alpha \colon  \Sigma^* \to  S$ such that
$(\alpha(w);f(w);(\alpha(w\sigma))_{\sigma \in \Sigma}) \in \T$
for all $w \in \Sigma^*$.
\end{remark}

\subsection{Graphical representation}
\begin{figure}[h!]
\centering
\subfigure[A general labeled transition bundle.]{
\begin{picture}(50,30)(-30,-15)
\gasset{Nw=7,Nh=7}
\node(S)(-15,0){$s$}
\node(SK)(15,12){$s_{k-1}$}
\node[Nframe=n,Nw=-1,Nh=-1](SD)(15,7){$\vdots$}
\node(SI)(15,0){$s_i$}
\node[Nframe=n,Nw=-1,Nh=-1](SDD)(15,-5){$\vdots$}
\node(S1)(15,-12){$s_0$}
\node[Nframe=n,Nw=-1,Nh=-1](SS)(0,0){}
\drawedge[AHnb=2,AHangle=20,AHLength=2,AHlength=0](S,SS){$a$}
\drawedge[curvedepth=3](SS,SK){$_{k-1}$}
\drawedge(SS,SI){$_i$}
\drawedge[curvedepth=-3](SS,S1){$_0$}
\end{picture}\label{FIGautomaton}
}\quad\quad
%%%%%%%%%%%%%%%%%%%%%%%%%%%%%%%%%%%%%%%%%%%%%%%%%%%%%%%%%%%%%%%%%%%%%%%
\subfigure[A labeled transition bundle of an unrestricted Rabin automaton in which $\Sigma = \{0,1\}$.]{
\begin{picture}(50,30)(-30,-15)
\gasset{Nw=7,Nh=7}
\node(S)(-15,0){$s$}
\node(S1)(15,5){$s_1$}
\node(S2)(15,-5){$s_0$}
\node[Nframe=n,Nw=-1,Nh=-1](SS)(0,0){}
\drawedge[AHnb=2,AHangle=20,AHLength=2,AHlength=0](S,SS){$a$}
\drawedge[curvedepth=3](SS,S1){}
\drawedge[curvedepth=-3,dash={1}0](SS,S2){}
\end{picture}\label{FIG2bundle}
}
\caption{Representations of a transition bundle.}
\label{FIGbundles}
\end{figure}
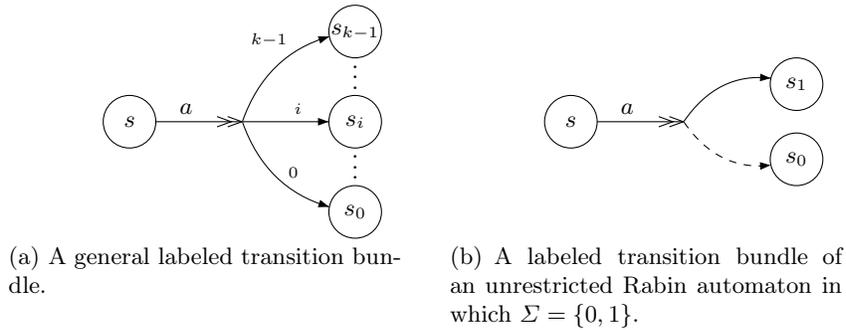

Let  $\vert \Sigma \vert = k$. We identify $\Sigma$ with the set  $\{0, 1, \dots, k-1\}$.
Hence, a transition bundle of an unrestricted Rabin automaton $\A = (S,\Sigma,A,\T)$ is a $(k+2)$-tuple $t = (s;a;s_0, \dots, s_{k-1})$ and it can be visualized as in Figure~\ref{FIGautomaton}. If $\vert \Sigma \vert = 2$ and $(s;a;s_0, s_1)$ is a transition bundle, we represent the edge from $s$ to $s_0$ by a broken line and the edge from $s$ to $s_1$ by a full line. This makes unnecessary to label the corresponding edges by $0$ and $1$, respectively (see Figure~\ref{FIG2bundle}).

\begin{figure}[h!]
\centering
%%%%%%%%%%%%%%%%%%%%%%%%%%%%%%%%%%%%%%%%%%%%%%%%%%%%%%%%%%%%%%%%%%%%%%%
\begin{picture}(0,10)(0,-5)
\gasset{Nw=7,Nh=7}
\node(A)(-15,0){$s_0$}
\node(B)(15,0){$s_1$}
\node[Nframe=n,Nw=-1,Nh=-1](AA)(-30,0){}
\node[Nframe=n,Nw=-1,Nh=-1](AB)(0,7){}
\node[Nframe=n,Nw=-1,Nh=-1](AB_)(0,-7){}
\node[Nframe=n,Nw=-1,Nh=-1](BB)(30,0){}
\drawedge[AHnb=2,AHnb=2,AHangle=20,AHLength=2,AHlength=0,ELside=r](A,AA){$0$}
\drawedge[curvedepth=5,dash={1}0](AA,A){}
\drawedge[curvedepth=8](AA,A){}
\drawedge[AHnb=2,AHnb=2,AHangle=20,AHLength=2,AHlength=0](A,AB){$0$}
\drawedge[curvedepth=1](AB,B){}
\drawedge[curvedepth=-1,dash={1}0](AB,B){}
\drawedge[AHnb=2,AHnb=2,AHangle=20,AHLength=2,AHlength=0](B,BB){$1$}
\drawedge[curvedepth=-5](BB,B){}
\drawedge[curvedepth=-8,dash={1}0](BB,B){}
\drawedge[AHnb=2,AHnb=2,AHangle=20,AHLength=2,AHlength=0,ELside=r](B,AB_){$1$}
\drawedge[curvedepth=1](AB_,A){}
\drawedge[curvedepth=-1,dash={1}0](AB_,A){}
\end{picture}
%%%%%%%%%%%%%%%%%%%%%%%%%%%%%%%%%%%%%%%%%%%%%%%%%%%%%%%%%%%%%%%%%%%%%%%
\caption{The unrestricted Rabin automaton accepting the tree shift of Example~\ref{e:monochromaticchildren}.}
\label{FIGmonochromaticchildren}
\end{figure}
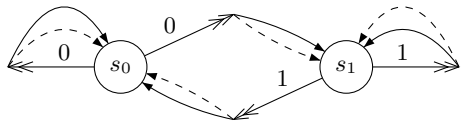

\begin{example}\label{e:AUTmonochromaticchildren}
Consider the unrestricted Rabin automaton $\A = (A,\Sigma,A,\T)$ where the bundle set is given by $$\T = \{(a;a;(a_\sigma)_{\sigma \in \Sigma}) \in A \times A \times A^\Sigma : a_\sigma = a_{\sigma'}
\textup{ for all } \sigma, \sigma' \in \Sigma\}.$$ We then have that $\Sh_\A$ is the
tree shift described in Example~\ref{e:monochromaticchildren}.
If $\vert \Sigma \vert = 2$ and $A = \{0,1\}$ the corresponding automaton is represented in Figure~\ref{FIGmonochromaticchildren}.
\end{example}

\subsection{Unrestricted Rabin automata and sofic shifts}
\begin{proposition}\label{p:XA sofic} Let $\A = (S,\Sigma,A,\T)$ be an unrestricted Rabin automaton.
Then $\Sh_\A$ is a sofic tree shift. Actually, up to a suitable extension of the alphabet $A$, there is an effective procedure to construct a tree shift of finite type $Y \subset A^{\Sigma^*}$ and a cellular automaton $\tau \colon Y \to A^{\Sigma^*}$ such that $\Sh_\A = \tau(Y)$.
\end{proposition}

Let $M \subset \Sigma^*$ be a nonempty subset and
$p \in A^M$ a pattern with support $M$. Given a word $w \in \Sigma^*$ we set $wM = \{wm : m \in M\} \subset \Sigma^*$
and denote by $^wp \in A^{wM}$ the pattern with support $wM$ defined by $(^wp)(wm) = p(m)$ for all $m \in M$.

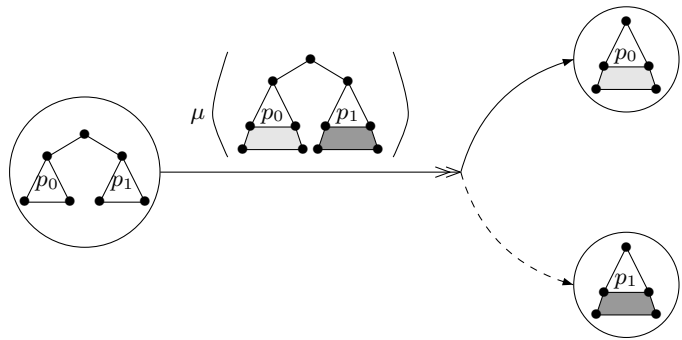
\begin{figure}[h!]
\centering
%%%%%%%%%%%%%%%%%%%%%%%%%%%%%%%%%%%%%%%%%%%%%%%%%%%%%%%%%%%%%%%%%%%%%%%
\begin{picture}(0,40)(0,-20)
\gasset{Nw=0,Nh=0,Nframe=n,ATnb=0,AHnb=0}
\node(dep)(-20,0){}
\node(arr)(20,0){}
\node(S0)(35,15){}
\node(S1)(35,-15){}
\drawedge[AHnb=2,AHnb=2,AHangle=20,AHLength=2,AHlength=0](dep,arr){}
\drawedge[AHnb=1,curvedepth=3](arr,S0){}
\drawedge[AHnb=1,curvedepth=-3,dash={1}0](arr,S1){}

\node(MU)(-15,7.5){$\mu$}
\drawpolygon[fillcolor=Gray,Nframe=y](-8,6)(-9,3)(-1,3)(-2,6)
\drawpolygon[fillcolor=Gray2,Nframe=y](2,6)(1,3)(9,3)(8,6)
\drawcurve(-11,16)(-13,7.5)(-11,2)
\drawcurve(11,16)(13,7.5)(11,2)
\node(1)(0,15){$\bullet$}
\node(2)(-5,12){$\bullet$}
\node(3)(5,12){$\bullet$}
\node(4)(-5,7.5){$p_0$}
\node(5)(5,7.5){$p_1$}
\node(6)(-8,6){$\bullet$}
\node(7)(-2,6){$\bullet$}
\node(8)(2,6){$\bullet$}
\node(9)(8,6){$\bullet$}
\node(10)(-9,3){$\bullet$}
\node(11)(-1,3){$\bullet$}
\node(12)(1,3){$\bullet$}
\node(13)(9,3){$\bullet$}
\drawedge(1,2){}
\drawedge(1,3){}
\drawedge(2,6){}
\drawedge(2,7){}
\drawedge(3,8){}
\drawedge(3,9){}

\node(A)(-30,5){$\bullet$}
\node(A0)(-35,2){$\bullet$}
\node(A1)(-25,2){$\bullet$}
\node(P0)(-35,-1.5){$p_0$}
\node(P1)(-25,-1.5){$p_1$}
\node(A00)(-38,-4){$\bullet$}
\node(A01)(-32,-4){$\bullet$}
\node(A10)(-28,-4){$\bullet$}
\node(A11)(-22,-4){$\bullet$}
\drawedge(A,A0){}
\drawedge(A,A1){}
\drawedge(A0,A00){}
\drawedge(A0,A01){}
\drawedge(A1,A10){}
\drawedge(A1,A11){}
\drawedge(A00,A01){}
\drawedge(A10,A11){}
\drawcircle[Nframe=y](-30,0,20)

\drawpolygon[fillcolor=Gray,Nframe=y](39,14)(38,11)(46,11)(45,14)
\node(AA0)(42,20){$\bullet$}
\node(PP0)(42,15.5){$p_0$}
\node(AA00)(39,14){$\bullet$}
\node(AA01)(45,14){$\bullet$}
\node(AA000)(38,11){$\bullet$}
\node(AA010)(46,11){$\bullet$}
\drawedge(AA0,AA00){}
\drawedge(AA0,AA01){}
\drawcircle[Nframe=y](42,15,14)

\drawpolygon[fillcolor=Gray2,Nframe=y](39,-16)(38,-19)(46,-19)(45,-16)
\node(AA1)(42,-10){$\bullet$}
\node(PP1)(42,-14.5){$p_1$}
\node(AA10)(39,-16){$\bullet$}
\node(AA11)(45,-16){$\bullet$}
\node(AA100)(38,-19){$\bullet$}
\node(AA110)(46,-19){$\bullet$}
\drawedge(AA1,AA10){}
\drawedge(AA1,AA11){}
\drawcircle[Nframe=y](42,-15,14)

\end{picture}
%%%%%%%%%%%%%%%%%%%%%%%%%%%%%%%%%%%%%%%%%%%%%%%%%%%%%%%%%%%%%%%%%%%%%%%
\caption{A transition bundle of $\A(\tau,M,X)$ when $\vert \Sigma\vert = 2$.}
\label{FIGdebruijn}
\end{figure}

\begin{definition}[Unrestricted Rabin automaton associated with a cellular automaton]\label{A(tau,M,X)}
{\rm Let $X \subset A^{\Sigma^*}$ be a tree shift of finite type and let $\tau \colon  X \to  A^{\Sigma^*}$ be a cellular automaton. Let $M = \Delta_n \subset \Sigma^*$ be a memory set for $\tau$ such that $n\geq 2$ and denote by $\mu \colon  A^M \to  A$ the corresponding local defining map. Fix $M' = \Delta_{n-1}$.
The \emph{unrestricted Rabin automaton $\A(\tau,M,X)$ associated with $\tau$} is defined by
$\A(\tau,M,X) = (X_{M'},\Sigma,A,\T),$ where $\T \subset X_{M'} \times A \times (X_{M'})^\Sigma$ consists of the bundles $(p;b;(p_\sigma)_{\sigma \in \Sigma})$ such that ({\it i.}) $p\vert_{\sigma {M'} \cap {M'}}$ equals $(^\sigma p_\sigma)\vert_{\sigma {M'} \cap {M'}}$ for all
$\sigma \in \Sigma$ (that is, $p(\sigma m) = p_\sigma(m)$ whenever $\sigma m \in \sigma {M'} \cap {M'})$;
({\it ii.}) the block $\bar p \colon  M \to  A$ coinciding with $p$ on $M'$ and with $^\sigma p_\sigma$ on $\sigma {M'}$ belongs to $X_M$ for all
$\sigma \in \Sigma$ (such a block $\bar p \in X_M$ is denoted by $\overline{(p;(p_\sigma)_{\sigma \in \Sigma})}$);
({\it iii.}) $b = \mu \left(\overline{(p;(p_\sigma)_{\sigma \in \Sigma})}\right)$.\\
A transition bundle of $\A(\tau,M,X)$ is illustrated in Figure~\ref{FIGdebruijn} for $\vert \Sigma\vert = 2$.}\end{definition}

%Note that since $M'$ is finite, so is $X_{M'}$ and therefore $\A(\tau,M,X)$ is also
%finite. In other words, $\A(\tau,M,X)$ is an unrestricted Rabin automaton.
%The following result generalizes a in the one-dimensional case.

\begin{proposition}\label{p:A(tau,M,X) accepts tau(X)} Let $X \subset A^{\Sigma^*}$ be a tree shift of finite type with memory $n-1$ and $\tau \colon  X \to  A^{\Sigma^*}$ be a cellular
automaton with memory set $\Delta_n$. Then $\Sh_{\A(\tau,\Delta_n,X)} = \tau(X)$.
\end{proposition}

\begin{remark}\label{r:transducer}
Conditions on $\Delta_n$ in Proposition~\ref{p:A(tau,M,X) accepts tau(X)} are not restrictive.
Proposition~\ref{p:A(tau,M,X) accepts tau(X)} says that $\A(\tau,\Delta_n,X)$ is a presentation of $\tau(X)$. In fact, we can actually show how to construct a pre-image of a configuration in $\Sh_{\A(\tau,\Delta_n,X)}$. This leads in particular to a presentation of $X$ as well.
\end{remark}

Proposition~\ref{p:XA sofic} and Proposition~\ref{p:A(tau,M,X) accepts tau(X)} imply the following result. The bottom-up version of it has been proved by B\'eal and Aubrun in~\cite{AubrunBeal12}.

\begin{corollary}\label{c:sofic iff accepted}
A tree shift $X \subset A^{\Sigma^*}$ is sofic if and only
if it is accepted by some unrestricted Rabin automaton.
\end{corollary}

\subsection{Deterministic and co-deterministic presentations}
An unrestricted Rabin automaton $\A =(S,\Sigma,A,\T)$ is \emph{deterministic} if, for each state $s\in S$, the transition bundles starting at $s$ carry different labels. Analogously, $\A$ is \emph{co-deterministic} if, for each sequence ${\bf s} \in S^\Sigma$, the transition bundles terminating at ${\bf s}$ (if there are any) carry different labels.

As stated below, for each unrestricted Rabin automaton $\A$ there exists a co-deterministic unrestricted Rabin automaton accepting the same shift.

\begin{theorem}[Subset construction]\label{t:subsetCONST}
Let $\A = (S,\Sigma,A,\T)$ be an unrestricted Rabin automaton. There exists a co-deterministic unrestricted Rabin automaton $\A_{\rm cod}$ such that $\Sh_\A = \Sh_{\A_{\rm cod}}$.
\end{theorem}
The statement of the above theorem fails to hold, in general, for deterministic unrestricted Rabin automata, as shown in the following counterexample.
%This equivalence between general unrestricted Rabin automata and the co-deterministic ones is used to show an algorithm to establish whether a cellular automaton defined on $A^{\Sigma^*}$ is surjective or not.

\begin{example}[A sofic shift not admitting a deterministic presentation]\label{e:countnondet}
Consider the tree shift $X$ presented in Example~\ref{e:monochromaticchildren}. A non-deterministic presentation of $X$ is given in Example~\ref{e:AUTmonochromaticchildren}.
Suppose that $X$ admits a deterministic presentation $\A = (S,\Sigma,A,\T)$. First observe that, in this case, each \emph{accessible} state (that is, each state that can be reached by a transition bundle), admits exactly one transition bundle starting at it. Thus for every accessible state $s \in S$ there exists exactly one configuration $f_s \in X$ accepted by a homomorphism $\alpha_s \colon \Sigma^* \to S$ \emph{starting} at $s$, that is, such that $\alpha_s(\varepsilon) = s$.
This implies that any state determines at most $|A|$ configurations (indeed, for a state $s$ that is not accessible, there are at most $|A|$ bundles that start at $s$ and all of these bundles end in accessible states). Therefore $\A$ accepts only finitely many different configurations, which contradicts the fact that $X$ is infinite.
\end{example}

\section{Full-tree-patterns and finite-tree automata}\label{section full-tree-patterns}
Recall that a \emph{$k$-ary rooted tree} is a rooted tree in which each vertex has at most $k$ children. A \emph{leaf} is a vertex without children. A \emph{full} $k$-ary rooted tree is a rooted tree in which every vertex other than the leaves has $k$ children. Hence $\Sigma^*$ is the full $k$-ary rooted tree with no leaves, where $k=\vert\Sigma\vert$. A \emph{subtree} of $\Sigma^*$ is a connected subgraph of $\Sigma^*$. We shall always suppose that a subtree of $\Sigma^*$ contains the root $\varepsilon$.
If $T\subset\Sigma^*$ is a subtree and $w \in T$,
we denote by $\Sigma_T(w)$ the set  $\{\sigma \in \Sigma : w\sigma \in T\}$. Hence $w\in T$ is a leaf if and only if $\Sigma_T(w) = \varnothing$.

Given a subtree $T$, we denote by $T^+$ the subtree $T \cup \{w\sigma : w \in T, \sigma \in \Sigma\}$. Notice that $T^+$ is always a full subtree. If $T$ is a full subtree, then $T^+$ is obtained by adding all the $k$ children of each leaf in $T$.

Notice that for each $n\geq1$ the set $\Delta_n$ is a full subtree whose leaves are the elements in $\Sigma^{n-1}$. Moreover, $\Delta_n^+ = \Delta_{n+1}$.

Finite full subtrees correspond to finite and complete prefix codes in~\cite{Aubrun11}.

%\begin{definition}{\rm
A pattern defined on a finite full subtree $T$ is called \emph{full-tree-pattern}.
The set of all full-tree-patterns is denoted by $\TT(A^{\Sigma^*})$. Given a shift $X \subset A^{\Sigma^*}$,
we denote by $\TT(X)$ the set of all full-tree-patterns of $X$ (that is, $\TT(X)= \bigcup_{T \subset \Sigma^*} X_T$, where the union ranges over all finite full subtrees $T$ of $\Sigma^*$).
%}\end{definition}

\begin{definition}[Sub-bundle]
{\rm Let $\A =(S,\Sigma,A,\T)$ be an unrestricted Rabin automaton. Let $M \subset \Sigma$ be a subset. A tuple $(s;a;(s_\sigma)_{\sigma \in M}) \in S \times A \times S^M$ is called a \emph{sub-bundle} of a transition bundle $(\bar s;\bar a;(\bar s_\sigma)_{\sigma \in \Sigma})\in \T$ provided $s = \bar s$, $a = \bar a$, and $s_{\sigma} = \bar s_\sigma$ for each $\sigma \in M$.
%The label $\lambda(\bar s;(\bar s_\sigma)_{\sigma \in \Sigma})$ is naturally associated with the sub-bundle $(s;(s_\sigma)_{\sigma \in M})$.
}\end{definition}

\begin{definition}\label{acceptanceSUBTREE}
{\rm Let $\A =(S,\Sigma,A,\T)$ be an unrestricted Rabin automaton. Let $T \subset \Sigma^*$ be a subtree and let $f \colon T \to A$ be a map. One
says that $f$ is \emph{accepted by $\A$} if there exists a map $\alpha \colon  T \to  S$ such that, for each $w \in T$,
$(\alpha(w);f(w);(\alpha(w\sigma))_{\sigma \in \Sigma_T(w)})$ is a sub-bundle of some $t \in \T.$
In this case we say that $f$ is accepted by $\A$ \emph{via} $\alpha$.
}\end{definition}

Note that, for a leaf $w \in T$, this latter acceptance condition reduces to saying that there exists a transition bundle starting at $\alpha(w)$ with label $f(w)$ (in fact, $\alpha$ is not defined on $w\sigma$ for any $\sigma\in \Sigma$).

\begin{proposition}\label{p:subtree} Let $\A =(S,\Sigma,A,\T)$ be an unrestricted Rabin automaton. Let $T \subset \Sigma^*$ be a subtree and suppose that $f \in A^T$ is accepted by $\A$. Then there exists a configuration $\bar f \in \Sh_\A$ such that $f = \bar f\vert_T$.
\end{proposition}

We have the following characterization of acceptance which immediately
results from Definition~\ref{acceptanceSUBTREE}.

\begin{proposition}\label{acceptanceSUBTREE2}
Let $\A =(S,\Sigma,A,\T)$ be an unrestricted Rabin automaton. Let $T \subset \Sigma^*$ be a finite full subtree. A full-tree-pattern $p \in A^T$ is accepted by $\A$ if and only if there exists a map $\alpha \colon  T^+ \to  S$ such that
%\begin{equation}\label{eqacceptanceSUBTREE2}
$(\alpha(w);p(w);(\alpha(w\sigma))_{\sigma \in \Sigma})\in \T$
%\end{equation}
for each $w \in T$.
\end{proposition}

By abuse of language, if this acceptance condition holds and there is no ambiguity, we say that the full-tree-pattern \emph{$p$ is accepted by $\A$ via $\alpha$}.
Obviously, Proposition~\ref{acceptanceSUBTREE2} applies whenever $T = \Delta_n$ for some $n \geq 1$ (recall that in this case $T^+ = \Delta_{n+1}$).

The following result follows from Proposition~\ref{p:subtree}.
\begin{corollary}
Let $\A =(S,\Sigma,A,\T)$ be an unrestricted Rabin automaton. Let $p\in \TT(A^{\Sigma^*})$ be a full-tree-pattern. Then $p\in \TT(\Sh_\A)$ if and only if $p$ is accepted by $\A$.
\end{corollary}

\begin{remark}\label{r:T(X) determines X}
The blocks of a subshift determine the subshift. In fact, given two subshifts $X,Y \subset A^{\Sigma^*}$, we have
$X = \Sh(\B(A^{\Sigma^*}) \setminus \B(X))$
so that
$X = Y$ if and only if $\B(X) = \B(Y)$. This fact obviously generalizes to full-tree-patters: $X = Y$ if and only if $\TT(X) = \TT(Y)$.
\end{remark}

\subsection{Finite-tree automata}\label{s:Finite-tree automata}
A \emph{finite-tree automaton} is an unrestricted Rabin automaton $\A = (S,\Sigma,A,\T)$ for which a subset $\mathcal{I} \subset S$ of \emph{initial states} and a state $F \in S$, called \emph{final state}, are specified. We shall denote it by $\A(\mathcal{I},F)$. We say that a full-tree-pattern $p \in A^T$ is \emph{accepted by $\A(\mathcal{I}, F)$} if there exists a map $\alpha \colon  T^+ \to  S$ such that
({\it i.}) $p$ is accepted by $\A$ via $\alpha$ (see Proposition~\ref{acceptanceSUBTREE2});
({\it ii.}) $\alpha(\varepsilon) \in \mathcal{I}$;
({\it iii.}) $\alpha(w) = F$ if $w \in T^+ \setminus T$.
We denote by $\TT(\A(\mathcal{I}, F))$ the set of all full-tree-patterns accepted by $\A(\mathcal{I}, F)$.
A set of full-tree-patterns is called \emph{recognizable} if it is of the form $\TT(\A(\mathcal{I}, F))$, for some finite-tree automaton $\A(\mathcal{I}, F)$.
A finite-tree automaton $\A(\mathcal{I}, F)$ is \emph{co-deterministic} if the unrestricted Rabin automaton $\A$ is co-deterministic.
\begin{remark}
As explained in Section~\ref{section full-tree-patterns}, we only consider essential unrestricted Rabin automata.
As far as finite-tree automata are concerned, we relax this assumption: each non final state is the source of some transition bundle, but no condition is required for the final state.
\end{remark}
An unrestricted Rabin automaton $\A = (S,\Sigma,A,\T)$ is called \emph{co-complete} if for each ${\bf s} \in S^\Sigma$ and $a \in A$, there exists a transition bundle in $\T$ labeled by $a$ and ending at~${\bf s}$. A finite-tree automaton $\A(\mathcal{I}, F)$ is \emph{co-complete} if the unrestricted Rabin automaton $\A$ is co-complete.

A slight adaptation in the proof of Theorem~\ref{t:subsetCONST} leads to the following result.
\begin{theorem}\label{t:subsetCONST2}
Let $\A$ be an unrestricted Rabin automaton. Then there is an effective procedure to construct a co-determini\-stic finite-tree automaton $\A_{\rm cod}(\mathcal{I}, F)$ such that
$\TT(\Sh_\A) = \TT(\A_{\rm cod}(\mathcal{I}, F)).$
\end{theorem}

The recognizable sets of full-tree-patterns form a class which is closed under complementation, as stated in the following theorem.

\begin{theorem}\label{t:complement}Let $\A(\mathcal{I}, F)$ be a co-deterministic finite-tree automaton. Then there exists a co-complete and co-deterministic finite-tree automaton $\A_\complement(\mathcal{I}_\complement, F_\complement)$ such that $\TT(A^{\Sigma^*}) \setminus \TT(\A(\mathcal{I}, F)) = \TT(\A_\complement(\mathcal{I}_\complement, F_\complement)).$
\end{theorem}

\begin{corollary}
\label{c:complement2}
Let $\A$ be an unrestricted Rabin automaton. Then there is an effective procedure to construct a co-complete and co-deterministic finite-tree automaton $\A_\complement(I,F)$ (with a single initial state) which accepts the complement of the set of all full-tree-patterns accepted by $\A$, in formul\ae, $\TT(\A_\complement(I,F)) = \TT(A^{\Sigma^*}) \setminus \TT(\Sh_\A)$.
\end{corollary}

\begin{corollary}\label{c:full}
Let $\A$ be an unrestricted Rabin automaton. Let $\A_\complement(I,F)$ be as in Corollary~\ref{c:complement2}. Then $\Sh_\A = A^{\Sigma^*}$ if and only if $\TT(\A_\complement(I, F)) = \varnothing$.
\end{corollary}

\subsubsection{The emptiness problem for finite-tree automata}\label{sec:emptiness}
The emptiness problem for an unrestricted Rabin automaton is trivial (every nonempty essential automaton accepts at least a configuration), but this argument does not apply to the case of finite-tree automata. In this section we present an effective procedure to establish the emptiness of recognizable set of full-tree-patterns.

First, we define
the \emph{height of a finite subtree} $T \subset \Sigma^*$ as the minimal $n \in \mathbb{N}$ such that $T \subset \Delta_n$. The \emph{height of a full-tree-pattern} $p \in A^T$ is the height of the (finite full) subtree $T$.

Let $\A(\mathcal{I}, F)$ be a finite-tree-automaton and let us show that there is an algorithm which establishes whether or not $\TT(\A(\mathcal{I}, F)) = \varnothing$.
Observe that $\TT(\A(\mathcal{I}, F))$ is nonempty if and only if it contains a pattern of height $\leq \vert S \vert$, where $S$ is the state set of $\A$ (we do not prove this fact in detail).
Since there are finitely many full-tree-patterns of height $\leq \vert S \vert$ one can effectively check whether or not they are accepted by $\A(\mathcal{I}, F)$.

Since in principle we have to check all possible maps $\alpha \colon \Delta_{\vert S \vert +1} \to  S$, this algorithm has exponential complexity in the size of $S$.

\subsubsection{An algorithm establishing whether two sofic shifts coincide}\label{par:equality}
 The \emph{join} of $\A_1 = (S_1,\Sigma,A,\T_1)$ and $\A_2 = (S_2,\Sigma,A,\T_2)$ is the unrestricted Rabin automaton $\A_1 * \A_2 = (S_1 \times S_2,\Sigma,A,\T_\times)$ where
$\left((s_1,s_2);a;(s'_\sigma,s''_\sigma)_{\sigma \in \Sigma}\right) \in \T_\times$ if and only if $\left(s_1;a;(s'_\sigma)_{\sigma \in \Sigma}\right) \in \T_1$ and~$\left(s_2;a;(s''_\sigma)_{\sigma \in \Sigma}\right) \in \T_2.$
Notice that $\Sh_{\A_1 * \A_2} = \Sh_{\A_1} \cap \Sh_{\A_2}$. Moreover, $\A_1 * \A_2$ is co-complete (respectively, co-deterministic), if $\A_1$ and $\A_2$ are co-complete (resp., co-deterministic).

We are now in position to describe our algorithm:
let $\A_1 = (S_1,\Sigma,A,\T_1)$ and $\A_2 = (S_2,\Sigma,A,\T_2)$ be two unrestricted Rabin automata. Note that, by Remark~\ref{r:T(X) determines X}, it suffices to establish whether or not \begin{equation}\label{whether-or-not}
\TT(\Sh_{\A_1}) \setminus \TT(\Sh_{\A_2}) = \varnothing = \TT(\Sh_{\A_2}) \setminus \TT(\Sh_{\A_1}).
\end{equation}

First construct the co-complete and co-deterministic finite-tree automata $\A_1'(I_1,F_1)$ and $\A_2'(I_2,F_2)$ as in Corollary~\ref{c:complement2}, associated with $\A_1$ and $\A_2$, respectively.
Consider the finite-tree automaton $(\A_1'*\A_2')(\mathcal{I}_1,F)$, where $\mathcal{I}_1 = (S_1\setminus\{I_1\}) \times \{I_2\}$ and $F = (F_1,F_2)$.
It can be seen that
$\TT((\A_1'*\A_2')(\mathcal{I}_1,F)) = \TT(\Sh_{\A_1}) \setminus \TT(\Sh_{\A_2})$.
Analogously, by defining $\mathcal{I}_2 = \{I_1\}\times (S_1\setminus\{I_2\})$ one has
$\TT((\A_1'*\A_2')(\mathcal{I}_2,F)) = \TT(\Sh_{\A_2}) \setminus \TT(\Sh_{\A_1})$.

Thus \eqref{whether-or-not} holds if and only if $\TT((\A_1'*\A_2')(\mathcal{I}_1,F)) \bigcup \TT((\A_1'*\A_2')(\mathcal{I}_2,F)) = \varnothing$.
An effective procedure to establish this latter equality is then provided by the solution to the emptiness problem.

\begin{remark}
The algorithm above has exponential complexity in the maximal size of the state sets of the unrestricted Rabin automata. A different procedure can be applied to the class of \emph{irreducible} unrestricted Rabin automata by using a minimization process. Actually, in~\cite{AubrunBeal10} it is shown that there exists a canonical minimal co-deterministic presentation of an irreducible sofic tree shift. Thus another possible decision algorithm consists in computing the minimal presentations of the two shifts and checking whether they coincide or not. In this case Theorem~\ref{t:subsetCONST} is needed while the procedure for the emptiness problem is not required. Hence this algorithm has in general an exponential complexity. The complexity can be reduced to be polynomial by only considering the class of co-deterministic irreducible tree shifts.
\end{remark}

\subsubsection{An algorithm establishing the surjectivity of cellular automata}
Observe first
that giving a sofic shift $X \subset A^{\Sigma^*}$ corresponds, equivalently, to giving a shift of finite type $Z \subset A^{\Sigma^*}$ and a surjective cellular automaton $\tau' \colon Z \to X$, or an unrestricted Rabin automaton $\A$ such that $X = \Sh_\A$. Propositions~\ref{p:XA sofic} and~\ref{p:A(tau,M,X) accepts tau(X)} provide an effective procedure to switch from one representation to the other.

Let $X, Y \subset A^{\Sigma^*}$ be two sofic shifts and $\tau \colon X \to Y$ a cellular automaton.
Let us show that it is decidable whether $\tau$ is surjective or not.
Let $Z \subset A^{\Sigma^*}$ and $\tau' \colon Z \to X$ as above.
Now the cellular automaton $\tau \colon X \to Y$ is surjective if and only if
the composite cellular automaton $\tau \circ \tau' \colon Z \to Y$ is surjective.
%Thus we can reduce to the case when $X$ is of finite type.
Let $n \in \mathbb{N}$ be large enough so that the cellular automaton $\tau \circ \tau'$ has memory set $\Delta_n$ and that $n-1$ is the memory of $Z$. By Proposition~\ref{p:A(tau,M,X) accepts tau(X)}, the unrestricted Rabin automaton $\A(\tau \circ \tau', \Delta_n, Z)$ having state set $Z_{n-1}$ is a presentation of $\tau(X)$. Then, it suffices to apply the algorithm in previous section to establish whether $Y=\tau(X)$.
%\qed
%\endproof
\begin{remark}
If $X = Y = A^{\Sigma^*}$, then the algorithm becomes much simpler.
Indeed, it can be proved by virtue of Corollary~\ref{c:full}, Corollary~\ref{c:complement2} and by using the emptiness algorithm.
\end{remark}

\bibliographystyle{siam}
\bibliography{biblio}

\end{document}